\documentclass[prd,preprint,groupedaddress,nofootinbib,showpacs,preprintnumbers,11pt]{revtex4}

\pdfoutput=1

\usepackage{epsfig}
\usepackage{amssymb}
\usepackage{amsbsy}
\usepackage{amsmath}
\usepackage{url}

\newcommand{\rf}[1]{(\ref{#1})}
\newcommand{\beq}{\begin{equation}}
\newcommand{\eeq}{\end{equation}}
\newcommand{\be}{\begin{equation}}
\newcommand{\ee}{\end{equation}}
\newcommand{\bea}{\begin{eqnarray}}
\newcommand{\eea}{\end{eqnarray}}
\newcommand{\eq}[1]{Eq.~(\ref{#1})}
\newcommand{\non}{\nonumber \\*}
\newcommand{\ie}{{i.e.}\ }
\newcommand{\bv}{\begin{verbatim}}
\newcommand{\ev}{\end{verbatim}}

\newcommand{\vp}{\varphi}

\newcommand{\e}{\,\mbox{e}}
\renewcommand{\d}{{\rm d}}
\renewcommand{\i}{{\rm i}}
\newcommand{\blambda}{\bar\lambda}
\newcommand{\brho}{\bar\rho}

\newcommand{\half}{{\textstyle \frac 12}}

\newcommand{\bz}{{\bar z}}

\newcommand{\eps}{\varepsilon}

\newcommand{\LA}{\left\langle}
\newcommand{\RA}{\right\rangle}

\def\fun#1#2{\lower3.6pt\vbox{\baselineskip0pt\lineskip.9pt
\ialign{$\mathsurround=0pt#1\hfil##\hfil$\crcr#2\crcr\sim\crcr}}}
\begin{document}

\preprint{ITEP--TH--05/21}
%\title{New anomaly that may cause difference between\\ 
%the Nambu-Goto and Polyakov strings}
\title{Private life of the Liouville field that causes\\ new anomalies in the Nambu-Goto string}
\author
{Yuri Makeenko}
\vspace*{2mm}
\affiliation{Institute of Theoretical and Experimental Physics,
B. Cheremushkinskaya 25, 117218 Moscow, Russia\\
\vspace*{1mm}
{\email: makeenko@itep.ru} }

%\today

\begin{abstract}
I consider higher-order terms of the Seeley expansion of the heat kernel, 
which for smooth metrics
are suppressed as inverse powers of the UV cutoff $\Lambda$, and demonstrate how they
result in an anomalous contribution to the string effective action after doing uncertainties 
$\Lambda^{-2}\times \Lambda^2$. For the Polyakov string these anomalies precisely 
reproduce at one loop the result of KPZ-DDK obtained for the Liouville theory by the 
conformal field theory technique. For
the Nambu-Goto string I find a deviation from this result which  shows that the two string
formulations may differ.
\end{abstract}


\maketitle

 \section{Introduction}

A challenging problem inherited from 1980's is that of existence of a quantum string
in four space-time dimensions. Studies of both the lattice discretization and the Polyakov
formulation of continuum bosonic string show that the theory is ill-defined unless the
dimension is lower than two.  Such a theory is associated 
with vast models of statistical mechanics whose continuum limits have beautiful description
in terms of two-dimensional gravity plus conformal matter.
Recently, a progress has been achieved in understanding of the no-go theorem for
the lattice strings (see Ref.~\cite{ADJ97} for a review). It has been shown~\cite{AM16}
that the continuum limit of a regularized  string should be taken in a very special way to 
guarantee stringy behavior. In this Paper I shall make an attempt to understand if
the results of applying the methods of conformal field theory to the string  could be
modified.

To be precise I mean the celebrated calculation of the string susceptibility index $\gamma_{\rm str}$
(also known as
the gravity anomalous dimension) by Knizhnik-Polyakov-Zamolodchikov~\cite{KPZ}  and
David~\cite{Dav88}, Distler-Kawai~\cite{DK89} often abbreviated as KPZ-DDK:
\be
\gamma_{\rm str}=(1-g)\left[\frac{d-25-\sqrt{(25-d)(1-d)}}{12}\right]+2
\label{ggg}
\ee
for a surface of genus $g$ embedded in $d$ Euclidean dimensions.
It is seen from \eq{ggg} that $\gamma_{\rm str}$ is not the real number 
for $1<d<25$ as it should. 
In a professional slang this was refer to as the $d=1$ barrier for the string existence.

The calculation of \rf{ggg} for the Polyakov string was based on 
the standard procedure of fixing the conformal gauge
for independent metric tensor $\rho_{ab}=\e^\vp \delta_{ab}$,
integrating over the embedded-space string coordinates $X^\mu$
and considering the resulting Liouville action for the remaining variable $\vp$ 
(for a good description of these steps see \cite{Pol87}).
DDK assumed that  the effective action, describing macroscopic distances, is again of the
Liouville type and applied the technique of conformal field theory to obtain \rf{ggg}.

My original motivation for this Paper was to reconsider the calculation for the Nambu-Goto string whose
action is just the area of the string world-sheet. Introducing an (imaginary) Lagrange 
multiplier $\lambda^{ab}$, the Nambu-Goto action can be written as
\be
S_{\rm NG}
=K_0 \!\int \sqrt{\det \partial_a X \cdot \partial_b X}=
K_0 \int \sqrt{\det  \rho} 
+\frac{K_0}2 \int \lambda^{ab} \left( \partial_a X \cdot \partial_bX -\rho_{ab}
\right), 
\label{aux}
\ee
where $K_0$ stands for the bare string tension. Path integrating over $X^\mu$ and 
$\lambda^{ab}$, we arrive at the {\em emergent}\/ action for $\vp$
\be
{\cal S} =\frac 1{16\pi b_0^2}\int \Big\{\partial_a \vp  \partial _a\vp 
+ \eps\e^{-\vp}\left[
(\partial ^2\vp)^2 +G\partial_a \vp\partial_ a\vp \partial^2 \vp
\right]+\mu^2 \e^\vp\Big\}+{\cal O}( \eps^2),
\label{NG1}
\ee
where $\eps$ is a UV cutoff at the world-sheet. The constant $G$ turns out to be different for
the Polyakov and Nambu-Goto strings ($G=0$ for the former and $G\neq 0$,  say $G=1$, for the latter).
For smooth $\vp$ we can drop the term with $\eps$ as $\eps\to0$, 
so  the difference between the two does 
not show up and we are left with the Liouville action. However, the additional
terms produce interactions for which $\eps$ plays the role of a coupling constant. Accounting
them perturbatively  results in divergences like powers of $\eps^{-1}$, so uncertainties of the 
type $\eps \times \eps^{-1}$ appear and have to be done.
This looks pretty much similar to the situation in certain non-renormalizable theories, 
{\it e.g.}\/ the
sigma model in three dimensions which becomes renormalizable if the coupling is $\sim \Lambda^{-1}$.

In order to study these uncertainties, I shall perform in this Paper an explicit computation
of the corresponding {\em effective}\/ action which would be again of the type of \rf{NG1}
but with some finite renormalization of the parameters. The computation is performed to
the first order of the expansion in $b_0^2$ and the resulting renormalization will be
some nonvanishing {\em universal}\/ numbers which seemingly 
do not depend on the form
of the higher-order terms in \rf{NG1} denoted as ${\cal O}(\eps^2)$. They come from
small distances $\sim \sqrt \eps$ and look like anomalies in quantum field theory.
This is why I say they are due to the private life of the Liouville field.

Our next goal in this paper would be to compare the beautiful intelligent results of KPZ-DDK,
obtained for the Polyakov string by
using conformal field theory technique, 
with the performed straightforward brute-force calculation
at one loop. Remarkably, I observe a precise agreement for $G=0$, \ie for the Polyakov string,
and a discrepancy for $G\neq0$ supposedly associated with the Nambu-Goto string.

This Paper is organized as follows.  In Sect.~\ref{s:comp} I present  the results obtained 
for the renormalization of the parameters in \rf{NG1} and compare them with KPZ-DDK.
In Sect.~\ref{s:Seel} I describe the setup for the calculation of the effective action
and perform it in Sect.~\ref{s:anom} for the Polyakov string at one loop. 
The universality of the results,
supporting the expectation we are dealing with string anomalies of a new kind, is
demonstrated in Sects.~\ref{s:univ} and \ref{s:renorm}, devoted to  the renormalization of the metric
tensor. Additional arguments in favor of the universality are
presented in Sect.~\ref{Jacob}, where the Jacobian associated with the transformation to
free fields is calculated. 
In Sect.~\ref{s:simp} I introduce a model which is a simplification to the Nambu-Goto string 
and yields the action~\rf{NG1} with $G=1$. The computations are  then
repeated for the $G\neq0$ case in Sect.~\ref{s:NGnew}.
In Sect.~\ref{s:conclu} I discuss the results.

\section{The results and comparison with KPZ-DDK\label{s:comp}}

Among several ways to derive KPZ-DDK I choose the original one~\cite{DK89} 
based on the background
(in)dependence which is beautifully described in \cite{ZZ}. 
If a Weyl factor  $\e^\vp$ is separated in the metric
\be
g_{ab} = \e^{\vp} \hat g_{ab},
\ee
then the curvature changes as
\be 
R = \e^{-\vp }\left( \hat R - \hat\Delta  \vp \right).
\ee
This produces the linear in $\vp$ term in the effective action 
\be
S=\frac1{8\pi b^2} \int \sqrt{\det{\hat g}}
\left[\frac 12 \hat g^{ab}\partial_a \vp \partial_b\vp+{q} \hat R \vp +\mu^2 \e^{\alpha\vp}\right],
\label{221}
\ee
where the differences of $b^2$ from $b_0^2$ and $q$ from $1$ are attributed to
the Jacobian for transition to the new field which  has the usual
Lebesgue measure in the path integral over $\vp$. I denote it also as $\vp$ because the
difference between the original variable with the nonlinear norm
\be
|| \delta \vp ||^2=\int \e^\vp (\delta \vp) ^2
\label{nonlin}
\ee
and the new one with the usual norm does not show up in the one-loop calculation.

For $\hat g_{ab}= \e^{\hat \vp} \delta_{ab}$ the linear term in \rf{221} can be compensated 
 by the shift 
 \be 
 \vp\to \vp- q \hat \vp .
\label{vshift}
\ee
The requirement for the effective action to be independent on $\hat \vp$ 
after this shift then results
in two equations
\begin{subequations}
\bea
-\frac 6{b_0^2}+1+\frac {6q^2}{b^2}&=&0,\qquad {b_0^2}=\frac{6}{26-d} 
\label{DDK1}\\
\label{DDK2}
\alpha q-\alpha^2 b^2&=&1 .
\eea
\label{DDK}
\end{subequations}
\!\!Equation~\rf{DDK1} implies the vanishing of the total central charge, while \eq{DDK2}
means that $\e^{\alpha\vp}$ is a primary field of conformal dimension 1. 
Both $\alpha$ and $b$ changes when  $\vp$ is multiplied by a constant but the product
$\alpha b$ does not change. There is no need to introduce both $\alpha$ and $b$
for the Liouville action \rf{221} but we shall need them both for the action  \rf{NG1} which
leads to interactions.

The solution to Eqs.~\rf{DDK} is
\be
\frac{1}{\alpha^2b^2}= \frac12 \left[\frac{1}{b^2_0} -\frac{13}{6}+
\sqrt{\left( \frac{1}{b^2_0} -\frac{25}6 \right)\left( \frac{1}{b^2_0} -\frac{1}6 \right)}\right]
\to \frac{1}{b^2_0} -\frac{13}{6} +{\cal O}(b_0^2).
\label{b2DDK}
\ee
This determines the string susceptibility index to be
\be
\gamma_{\rm str}=(1-g)\frac q{\alpha b^2} +2,\qquad
\frac q{\alpha b^2} = \frac{1}{b^2_0} -\frac{7}{6} +{\cal O}(b_0^2),
\label{gstr}
\ee
yielding \rf{ggg}.
This formula simply follows from a uniform dilatation of space, which means adding a constant
value to $\alpha\vp$. Then the second term in the brackets in 
\rf{221} becomes the topological Gauss-Bonett term, explaining why
the Euler characteristic $2-2g$ has appeared in \rf{gstr}.

One of the results of this Paper is an explicit computation of  $b^2$ and $\alpha$ 
 to the leading order of the expansion in $b_0^2$ (\ie at one loop). These are given 
 for the Polyakov string by Eqs.~\rf{bbbb} and \rf{aaaa} below:
\be
\frac1{b^2}=\frac1{b_0^2} -\frac 16+2 +{\cal O}(b_0^2)
\label{bbbb0}
\ee
and
\be
\alpha=1+2b_0^2+{\cal O}(b_0^4).
\label{aaaa0}
\ee 
Combining \rf{bbbb0} and \rf{aaaa0} we remarkably obtain
\be
\frac1{\alpha^2 b^2}= \frac {1}{b_0^2} -\frac{13}6 +{\cal O}(b_0^2),
\label{remark}
\ee
exactly reproducing \eq{b2DDK} of DDK to the given order.

Alternatively, for the action \rf{NG1} with $G\neq0$ we find that $\alpha$ does not change while
\be
\frac1{\alpha^2 b^2}= \frac {1}{b_0^2} -\frac{13}6 +2G+{\cal O}(b_0^2).
\label{remark2}
\ee
This gives a clear discrepancy from KPZ-DDK for the simplified model 
 introduced in Sect.~\ref{s:simp} which is associated with the Nambu-Goto 
string and where $G=1$.

\section{Seeley et al. expansion\label{s:Seel}}

Let us begin by recalling the structure of the Seeley expansion of the heat kernel
\be
\LA \omega \Big| \e^{a^2 \Delta }
\Big| \omega \RA = \frac 1{4\pi a^2}
%\Lambda^2   %%%\frac{\sqrt{\det{\rho}}}{\sqrt{\det{\lambda}}}
 +\frac1{24\pi} {R(\omega)}+\frac{a^2}{120\pi}
\left( \Delta R(\omega)+\frac 12 R^2(\omega)
\right)+
\ldots\,,
\qquad a^2=\frac1{4\pi \Lambda^2}
%%%+{\cal O}(\tau^0)
\label{Seeley}
\ee
which is customly  used to integrate over $d$ target-space coordinates
$X^\mu$ in the Polyakov string formulation.
Here $\Delta$ denotes the two-dimensional Laplacian and
$a$ is related to the UV cutoff $\Lambda$ as shown in \rf{Seeley}. 
%%It thus becomes an expansion in $1/\Lambda^2$. 
Equation~\rf{Seeley} was originally derived%
\footnote{We use the definition of the curvature $R=-4\e^{-\vp} \partial \bar\partial \vp$
in the conformal gauge with
an opposite sign to that in \cite{deWitt,Gil75}.} in \cite{deWitt,Gil75}
for closed curved spaces
and generalized to the spaces with  boundaries in \cite{DOP82,Alv83}. 

Dropping higher terms in $a^2$, which are denoted in the expansion \rf{Seeley} by $\ldots$,
results  in the conformal gauge 
$\rho_{ab}=\brho \e^\vp \delta_{ab}$   in the contribution to 
the effective action of $\vp$
\be
{\cal S}_X= \int \left[-\frac d{24 \pi} \partial \vp \bar\partial \vp+ \frac {d a^2}{30\pi \brho}
\e^{-\vp} (\partial \bar\partial \vp)^2 \right], 
\label{LX}
\ee
where $\brho=1$ for the classical string ground state, but $\brho$
has a nontrivial value for the
mean-field ground state~\cite{AM16} which turns out to be stable for $2<d<26$.
Accounting for ghosts, we get finally
\be
{\cal S}= \frac 1{4 \pi b^2_0}\int \left[ \partial \vp \bar\partial \vp+ 4\eps
\e^{-\vp} (\partial \bar\partial \vp)^2 \right], \qquad b^2_0=\frac 6{26-d},
\label{S}
\ee
where $\eps \propto a^2/\brho$ and depends in general on the regularization applied.
We have dropped in \rf{S}
the exponential term because its coefficient (denoted in \eq{NG1} by $\mu^2$)
vanishes~\cite{AM16} for the stable ground state, minimizing the action  in $2<d<26$.

The second term in \eq{S} is usually omitted for smooth metrics when $R\ll \Lambda^2$.
However, this term not only changes the propagator but also produces self-interaction 
of $\vp$ which results in diagrams with
quadratic divergences  like powers of $\Lambda^2$. % at the one-loop order. 
If we treat $\eps$ in \rf{S} as a coupling constant, then terms like $\eps \times \Lambda^2$
appear which are $\sim 1$ for $\eps\sim 1/\Lambda^2$.
We shall do below this uncertainty at the one-loop order of the ``semiclassical'' expansion 
in $b^2_0$ about the ground state.

Each of the  two terms in \rf{S} is invariant under 
the usual conformal transformation
\be
\delta \vp(z,\bar z) = \xi'(z) +\xi(z) \partial \vp(z,\bar z)+{\rm h.c.} .
\label{cta0}
\ee
 For this reason we expect that the action
\rf{S} will have conformal structures  like KPZ-DDK.
We shall explicitly show this below at the one-loop order of the expansion in $b_0^2$.

In the computations we shall use the Pauli-Villars regularization, adding to \rf{S}
the action of the regulators
\be
{\cal S}_{\rm reg.}= \frac 1{4 \pi b_0^2}\int \left\{ 
\sum_{i=1}^2 \left[ \partial Y_i \bar\partial Y_i +
\textstyle{\frac 14}M^2 Y_i^2+ 4\eps
\e^{-\vp} (\partial \bar\partial Y_i)^2 \right]+
 \left[ \partial Z\bar\partial Z +\half M^2 Z^2+ 4\eps
\e^{-\vp} (\partial \bar\partial Z)^2 \right]\right\}, 
\label{Sreg}
\ee
where the regulator fields $Y_i$ ($i=1,2$) and $Z$ have, respectively, ghost and 
usual statistics
 and masses squared $M^2$ and $2M^2$ as is outlined in \cite{AM17c}. 
The total action we shall use 
in calculations is thus
\be
{\cal S}_{\rm tot.}={\cal S}+{\cal S}_{\rm reg.}
\label{Stot}
\ee
It will regularize all divergences that appear.

\section{Effective action at one loop\label{s:anom}}

\subsection{Divergent part}

\begin{figure}
\includegraphics[width=6cm]{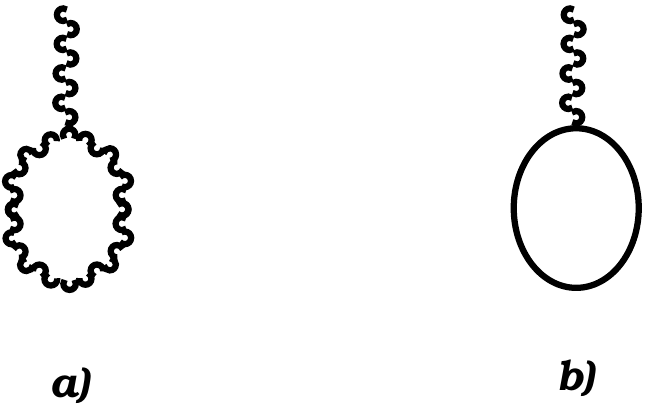} 
\caption{One-loop tadpole diagrams contributing to the effective action to order $\vp$. The wavy lines represent $\vp$, while the solid line represents  the regulator fields.}
\label{fi:tad}
\end{figure}

Let us consider renormalization of the action \rf{S}, distinguishing fast (quantum) and slow (classical) fluctuations of $\vp$ and computing an effective action for the latter.
The quadratic divergence of the effective action comes to order $\vp$ from the
tadpole diagrams in Fig.~\ref{fi:tad}.
To compare with the previously known results~\cite{Pol86,Klei86,OY86,Bra86}
for the rigid string, let us start from the diagram in Fig.~\ref{fi:tad}b, 
where the solid line represents  the regulators $Y_i^\mu$ and $Z^\mu$ ($\mu=1,\ldots,d$).
We have $d=1$ for our problem, but let us consider an arbitrary $d$.
Analogously, let us associate the factor $d$ with the diagram in Fig.~\ref{fi:tad}a like if
we have $d$ fields $\vp$. This is the same as if we have the contribution to the tadpole in 
Fig.~\ref{fi:tad}b from the field $X^\mu$ with the action like that for $Z^\mu$ in \eq{Sreg}
but zero mass.
The right result for the sum of the diagrams in Fig.~\ref{fi:tad}
would be when  $d=1$ in the formulas below, but 
for the sake of the comparison let us temporary keep $d$ arbitrary.

We then find for the contribution of the diagrams in Fig.~\ref{fi:tad} to the effective action
\bea
{\rm Fig.~\ref{fi:tad}}&= & -\frac d2 \int \frac{\d^2 p}{(2\pi)^2}\vp(p) \,
\int \frac{\d^2 k}{(2\pi)^2}\left[ \frac{\eps k^4}{(k^2+\eps k^4)}
-2\frac{\eps k^4-M^2}{(k^2+M^2+\eps k^4)}
+\frac{\eps k^4-2M^2}{(k^2+2M^2+\eps k^4)}
\right] \non &=&
=- \frac d2 \int \frac{\d^2 p}{(2\pi)^2}\vp(p) \Lambda^2
\label{tad1}
\eea
with
\be
\Lambda^2=\frac{1}{8\pi \eps} \left[%%-\log\frac{M^2\eps}2+
4\sqrt{4 M^2\eps-1} 
\arctan\left(\sqrt{4 M^2\eps-1}\right) -2\sqrt{8 M^2\eps-1}
\arctan \left(\sqrt{8 M^2\eps-1} \right)-\log\frac{M^2\eps}2\right].
\label{La2}
\ee
If the quartic in the derivatives term in the action vanishes which means $\eps\to0$, we have
\be
 \Lambda^2\stackrel{\eps\to0}\to
\frac{M^2}{2\pi} \log 2,
\ee
reproducing the correct result for the Polyakov string.
Alternatively, for finite $\eps$ and $M\to\infty$ we have
\be
\Lambda^2 \stackrel {M\to \infty}\to \frac 1{4}
\left[(2-\sqrt 2) \frac{M}{\sqrt\eps} -\frac 1{2\pi} \log (M^2\eps)
\right]
\label{alt}
\ee
with the log familiar from the rigid string~\cite{Pol86,Klei86}.

\begin{figure}
\includegraphics[width=13cm]{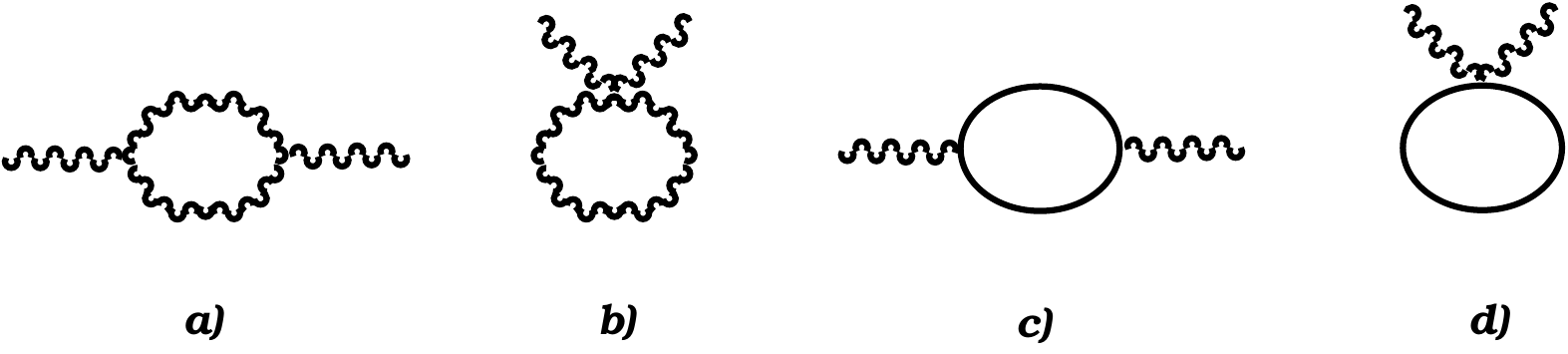} 
\caption{One-loop diagrams contributing to the effective action to order $\vp^2$. 
The wavy lines represent $\vp$, while the solid lines represent  the regulator fields.}
\label{fi:tad2}
\end{figure}

Finite terms in the effective action come from the diagrams in Fig.~\ref{fi:tad2},
which we compute again for an arbitrary $d$.
Let us begin with the divergent parts of the diagrams in Fig.~\ref{fi:tad2}. We have
\bea
\lefteqn{{\rm Fig.~\ref{fi:tad2}a}\Big|_{\rm div}+
{\rm Fig.~\ref{fi:tad2}c}\Big|_{\rm div}}\non &&=  -\frac 12 \times \frac d2\int \vp^2
%%\frac{\d^2 p}{(2\pi)^2}\vp(-p)\vp(p) \,
\int \frac{\d^2 k}{(2\pi)^2}\left[ \frac{(\eps k^4)^2}{(k^2+\eps k^4)^2}
-2\frac{(\eps k^4-M^2)^2}{(k^2+M^2+\eps k^4)^2}
+\frac{(\eps k^4-2M^2)^2}{(k^2+2M^2+\eps k^4)^2}
\right] 
\label{tad2c}
\eea
and
\bea
\lefteqn{{\rm Fig.~\ref{fi:tad2}b}\Big|_{\rm div}+
{\rm Fig.~\ref{fi:tad2}d}\Big|_{\rm div}}\non&&=  \frac12 \times
\frac d2 \int \vp^2
%%\frac{\d^2 p}{(2\pi)^2}\vp^2(p) \,
\int \frac{\d^2 k}{(2\pi)^2}\left[ \frac{\eps k^4}{(k^2+\eps k^4)}
-2\frac{\eps k^4+M^2}{(k^2+M^2+\eps k^4)}
+\frac{\eps k^4+2M^2}{(k^2+2M^2+\eps k^4)}
\right].
\label{tad2d}
\eea
The results for both \rf{tad2c} and \rf{tad2d} look ugly but their sum is rather simple
\be
{\rm Fig.~\ref{fi:tad2}a}\Big|_{\rm div}+
{\rm Fig.~\ref{fi:tad2}b}\Big|_{\rm div}+
{\rm Fig.~\ref{fi:tad2}c}\Big|_{\rm div}+{\rm Fig.~\ref{fi:tad2}d}\Big|_{\rm div}
=- \frac d4 \Lambda^2 \int \vp^2
%%%\frac{\d^2 p}{(2\pi)^2}\vp(-p) \vp(p)
\ee
with $\Lambda^2$ given by \eq{La2},
reproducing together with \rf{tad1} the expansion of $\e^\vp$.

\subsection{Finite part}

To compute the finite part, let us start from the diagram Fig.~\ref{fi:tad2}c 
adding again a part of the diagram in  Fig.~\ref{fi:tad2}a whose contribution is the same as
if we have the contribution to the diagram Fig.~\ref{fi:tad2}c from $d$ massless fields $X^\mu$.
It is given by the first term in the curly brackets.
This remarkably reproduces the conformal anomaly
\bea
{\rm Fig.~\ref{fi:tad2}c} =-\frac 12 \times \frac d2
\int\frac{\d^2 p}{(2\pi)^2}\vp(-p)\vp(p) \,
\int \frac{\d^2 k}{(2\pi)^2}\left\{ \frac{[\eps k^2(k-p)^2]^2}
{(k^2+\eps k^4)[(k-p)^2+\eps(k-p)^4]} \right. \non\left.
-2\frac{[\eps  k^2(k-p)^2-M^2]^2}{(k^2+M^2+\eps k^4)[(k-p)^2+M^2+\eps(k-p)^4]} 
\right. \non \left.
+\frac{[\eps k^2(k-p)^2-2M^2]^2}{(k^2+2M^2+\eps k^4)[(k-p)^2+2M^2+\eps(k-p)^4]}
\right\} \nonumber  \\
={\rm Fig.~\ref{fi:tad2}c}\Big|_{\rm div} -\frac d{96\pi} 
\int\frac{\d^2 p}{(2\pi)^2} p^2\vp(-p)\vp(p) +{\cal O}(M^{-2}) . \hspace*{3.4cm}
\label{tad2cfin}
\eea

It is not yet the whole story because the diagrams in Figs.~\ref{fi:tad2}a, b
may also have finite parts. A~part of the diagram in Fig.~\ref{fi:tad2}a is already taking 
into account by
the above formulas with $d=1$. The additional parts are due to the non-quadratic dependence
of the action on $\vp$ and do not involve the regulators.
Additionally we have
\be
{\rm Fig.~\ref{fi:tad2}a} \Big|_{\rm add} 
=  -\frac 12\int (\partial_a\vp)^2
%%\frac{\d^2 p}{(2\pi)^2}\vp^2(p) \,
\int \frac{\d^2 k}{(2\pi)^2}\frac{\eps k^2 \eps k^4}{(k^2+\eps k^4)^2}
\label{pp1}
\ee
and
\be
{\rm Fig.~\ref{fi:tad2}b} \Big|_{\rm add} 
= \frac 12  \int (\partial_a\vp)^2
%%\frac{\d^2 p}{(2\pi)^2}\vp^2(p) \,
\int \frac{\d^2 k}{(2\pi)^2}\frac{\eps k^2}{(k^2+\eps k^4)}.
\label{pp2}
\ee
Both \rf{pp1} and \rf{pp2} logarithmically diverge but their sum is finite and equals 
\be
{\rm Fig.~\ref{fi:tad2}a\Big|_{\rm add}+Fig.~\ref{fi:tad2}b} \Big|_{\rm add} = \frac1{8\pi}\int (\partial_a\vp)^2.\label{ppfin}
\ee
It has the same structure as the conformal anomaly \rf{tad2cfin} but an opposite sign
and also contributes to the effective action. 
%%This extends the result \cite{Klei86} for the rigid
%%string where this additional contribution was missing.
One again, it has appeared as a result of doing uncertainty $\eps \times \eps^{-1}$ with
$\eps \sim \Lambda^{-2}\to 0$.
The cancellation of the logs which would otherwise spoil conformal invariance at one loop is a 
manifestation of the theorem formulated in \cite{Mak18} that is based on the quadratic form
of the effective action governing smooth fluctuations essential in the infrared.

Summing \rf{tad2cfin} with $d=1$ and \rf{ppfin}, we find
\be
\frac1{b^2}=\frac1{b_0^2} -\frac 16+2 +{\cal O}(b_0^2).
\label{bbbb}
\ee
This formula shows that the ``bare'' constant $b_0^2$ undergoes a finite renormalization because
of the interaction implied by the action~\rf{S}.

\section{The universality \label{s:univ}}

Let us discuss the universality of the obtained results in the sense of their independence of the
exact
form of the action. We substitute \rf{S} by a more general action, adding the higher terms,
\be
{\cal S}= -\frac{1}{16 \pi b_0^2} \int \sqrt{g} \vp \Delta F(-\eps \Delta) \vp
\label{Sgen}
\ee
and verify whether the results will not depend on the choice of the function
\be
F(\eps x)=1+\sum_{n\geq1} f_n \eps^n x^{n},\qquad f_1=1,\quad F(\infty)=\infty.
\ee

The next complicated case is $f_2\neq 0$ and $f_n=0$ for $n\geq 3$. The action \rf{Sgen}
to the quartic order in $\vp$ then reads
\bea
{\cal S}&=& \frac{1}{16 \pi b_0^2} \int \big[\vp \left(-\partial^2 
+\eps \partial^2 \e^{-\vp} \partial^2 -
f_2 \eps^2 \partial^2 \e^{-\vp} \partial^2\e^{-\vp} \partial^2\right)\vp \big]\non
&= &\frac{1}{16 \pi b_0^2} \int \big[\vp \left(-\partial^2 +\eps \partial^4 -
f_2 \eps^2 \partial^6 \right)\vp -\eps (\vp -\half \vp^2) (\partial^2 \vp)^2 
+2 f_2 \eps^2  (\vp -\half \vp^2)  \partial^2 \vp\partial^4 \vp \non && \hspace*{1.67cm}
-f_2 \eps^2 \vp^2 \partial^2 \vp \partial^4 \vp
+f_2 \eps^2 \partial_a \vp \partial_a\vp (\partial^2 \vp)^2 \big]
+{\cal O}(\vp^5)
\label{Sgen2}
\eea
which generates three- and four-point vertices. For the regulators we have analogously
\bea
{\cal S}_{\rm Reg}&=& \frac{1}{16 \pi b_0^2} \int \big[Y \left(-\partial^2 +M^2\e^\vp
+\eps \partial^2 \e^{-\vp} \partial^2 -
f_2 \eps^2 \partial^2 \e^{-\vp} \partial^2\e^{-\vp} \partial^2\right) Y \big]\non
&= &\frac{1}{16 \pi b_0^2} \int \big[Y \left(-\partial^2 +M^2+\eps \partial^4 -
f_2 \eps^2 \partial^6 \right)Y +M^2(\vp-\half \vp^2)Y^2
-\eps (\vp -\half \vp^2) (\partial^2 Y)^2 \non && \hspace*{1.67cm}
+2 f_2 \eps^2  (\vp -\half \vp^2)  \partial^2 Y\partial^4 Y 
-f_2 \eps^2 \vp^2 \partial^2 Y \partial^4 Y
+f_2 \eps^2 \partial_a \vp \partial_a\vp (\partial^2 Y)^2 \big]
+{\cal O}(\vp^3). \non &&
\label{Sgen2reg}
\eea
Higher orders in $\vp$ in these actions are again not
essential at one loop.

The parts of the diagrams in Fig.~\ref{fi:tad2}c (with the inclusion a part of 
the diagram in Fig.~\ref{fi:tad2}a as before)
and Fig.~\ref{fi:tad2}d which 
reproduce the conformal anomaly are
\bea
\lefteqn{{\rm Fig.~\ref{fi:tad2}c}\Big|_{\rm fin}
=-\frac 12 \times \frac d2
\int\frac{\d^2 p}{(2\pi)^2}\vp(-p)\vp(p) }  \non&& \times
\int \frac{\d^2 k}{(2\pi)^2}\left\{ \frac{[\eps k^2(k-p)^2+f_2 \eps^2  
(k^2(k-p)^4+ k^4(k-p)^2)]^2}
{(k^2+\eps k^4+f_2 \eps^2 k^6)[(k-p)^2+\eps(k-p)^4+f_2 \eps^2 (k-p)^6]} \right. \non && \hspace*{2cm}\left.
-\frac{[\eps  k^2(k-p)^2-M^2+f_2 \eps^2  
(k^2(k-p)^4+ k^4(k-p)^2)]^2}{(k^2+M^2+\eps k^4+f_2\eps^2 k^6)
[(k-p)^2+M^2+\eps(k-p)^4+f_2 \eps^2 (k-p)^6]}
\right\} 
\label{tad2cfingen}
\eea
and
\be
{\rm Fig.~\ref{fi:tad2}d}\Big|_{\rm fin}=\frac d2
\int\frac{\d^2 p}{(2\pi)^2}p^2\vp(-p)\vp(p) 
\int \frac{\d^2 k}{(2\pi)^2} \left[\frac{f_2\eps^2 k^4}{(k^2+\eps k^4+f_2 \eps^2 k^6)}
-\frac{f_2\eps^2 k^4}{(k^2+M^2+\eps k^4+f_2 \eps^2 k^6)} \right].
\label{tad2dfingend}
\ee
We have explicitly written here only one regulator with mass $M$ because the anomaly 
does not depend on the mass. The two additional terms in \eq{Sreg} were needed only to
regularize the divergent part. The finite part of the sum of \rf{tad2cfingen} and
 \rf{tad2dfingend} gives precisely the conformal anomaly
\be
{\rm Fig.~\ref{fi:tad2}c} \Big|_{\rm fin}+{\rm Fig.~\ref{fi:tad2}d} \Big|_{\rm fin}
=-\frac d{96\pi} 
\int\frac{\d^2 p}{(2\pi)^2} p^2\vp(-p)\vp(p) +{\cal O}(M^{-2}) 
\ee
for any  $\eps$ and $f_2$.

There is still a bunch 	of additional diagrams where $\vp$ pairs with $\vp$ standing in place
of $Y$. We have
\be
{\rm Fig.~\ref{fi:tad2}a}\Big|_{\rm add}=-\frac 12
\int\frac{\d^2 p}{(2\pi)^2}p^2\vp(-p)\vp(p) 
\int \frac{\d^2 k}{(2\pi)^2} 
\frac{(\eps^2 k^6 +3 f_2\eps^3 k^8+2 f_2^2 \eps^4 k^{10})}{(k^2+\eps k^4+f_2 \eps^2 k^6)^2}
\label{tad2dfingc}
\ee
and
\be
{\rm Fig.~\ref{fi:tad2}b}\Big|_{\rm add}=\frac 12
\int\frac{\d^2 p}{(2\pi)^2}p^2\vp(-p)\vp(p) 
\int \frac{\d^2 k}{(2\pi)^2} 
\frac{(\eps k^2 +2f_2\eps^2 k^4)}{(k^2+\eps k^4+f_2 \eps^2 k^6)}.
\label{tad2dfingd}
\ee
The sum of \rf{tad2dfingc} and \rf{tad2dfingd} does not depend on $\eps$ and $f_2$ 
and is equal to \rf{ppfin}. This is therefore a strong argument that we are dealing indeed
with a new kind of anomalies. Hopefully, the derivation can be extended to an arbitrary
function $F$ in \eq{Sgen}. An illustration of how it may work is \eq{vvvgen} below.

\section{Field redefinition and Jacobians\label{Jacob}}

Let us present yet another derivation of the above result by computing the determinants associated with field redefinition. 
The action \rf{S} can be deduced from the free action given by the first term in \rf{S} by
virtue of the field redefinition
\be
\vp\to\vp'= \sqrt{1-\eps \Delta }\,\vp .
\label{cheall}
\ee
However,  a determinant of a nontrivial operator is
produced by the measure for $\vp$ while changing \rf{cheall}, 
so we have to investigate how this procedure may reproduce the above results.
%%or may not be equivalent with the quadratic action by DDK.

To regularize  we make the change of the regulators analogously to \rf{cheall}
\be
Y_i\to Y_i'= \sqrt{1+\eps M^2 -\eps \Delta }\,Y_i,\qquad 
Z\to Z'= \sqrt{1+2\eps M^2-\eps \Delta }\,Z .
\label{cheallR}
\ee
This produces the determinants
\be
\frac{{\cal D }Y_i}{{\cal D} Y'_i} =\det\mbox{}^{1/2} \left(1+\eps M^2-\eps \Delta \right),
\qquad
\frac{{\cal D }Z}{{\cal D} Z'} =\det\mbox{}^{-1/2} \left(1+2\eps M^2-\eps \Delta \right).
\ee
The positive power of the determinant for $Y_i$ is because they are Grassmann variables.

It is more complicated with the Jacobian associated with the change \rf{cheall} 
because it  depends on $\vp$ non-linearly and we need the inverse function $\vp(\vp')$. 
We can simply write 
\be
\vp=\frac{1}{\sqrt{1-\eps \Delta}} \vp'
\label{trial}
\ee
which gives the correct term of the order $\eps$, but may differ from exact $\vp(\vp')$ by
higher orders in $\eps$. But if we believe in the universality, we may think that the trial
function~\rf{trial} will give the right result at least at one loop.
To the leading order in $\eps$ we then write
\be
\frac{{\cal D }\vp}{{\cal D} \vp'} =
\frac{\det  \sqrt{1-\eps \Delta }}{\det{\left(1-\eps \Delta +\eps R/2\right)}}.
\ee
For the total Jacobian we thus write
\be
\frac{{\cal D }\vp}{{\cal D} \vp'} \prod_{i=1}^2\frac{{\cal D }Y_i}{{\cal D} Y'_i} 
\frac{{\cal D }Z}{{\cal D} Z'} =R_1^{-1} R_2^{-1},
\label{ratiodet}
\ee
where 
\be
R_1=\frac{\det{\left(1-\eps \Delta-\eps R/2 \right)}}
{\det{\left(1-\eps \Delta \right)}}=
\frac{\det{\left(\e^{\vp'}-\eps \partial^2+\eps \partial^2\vp'/2 \right)}}
{\det{\left(\e^{\vp'}-\eps \partial^2 \right)}}
\label{R1}
\ee
and
\be
R_2=\frac{\det\sqrt{1-\eps \Delta }
\det\sqrt{1+2\eps M^2-\eps \Delta}}{\det \left(1+\eps M^2-\eps \Delta \right)}=
\frac{\det^{1/2}{\left(\e^{\vp'}-\eps \partial^2 \right)}
\det^{1/2}{\left(\e^{\vp'}+2\eps M^2-\eps \partial^2\right)}}
{\det \left(\e^{\vp'}+\eps M^2-\eps \partial^2 \right)}.
\label{R2}
\ee

As far as the ratio $R_2$ is concerned, it has only (regularized)
divergent part and no finite part because
the conformal anomalies mutually cancel in the ratio. On the contrary, the divergent parts
of the determinants cancel in the ratio $R_1$ as well as the usual conformal anomaly does.
Naively we expect that $R_1$ tends to 1 when $\eps\to0$, but
 to order $\vp'^2$ the result  is given by the diagram in Fig.~\ref{fi:tad2}c and reads
\be
\log R_1=\frac 12
\int \partial_a \vp' \partial_a \vp' \int \frac{\d^2 k}{(2\pi)^2} 
\frac{\eps k^2} {\left(1+\eps k^2\right)^{2}} = \frac 1{8\pi}\int \partial_a \vp' \partial_a \vp'
\label{aa1}
\ee
which coincides with \rf{ppfin}.

The computation of $R_1$ can be easily extended to all orders in $\vp'$, keeping in mind
that we need only the first order of the expansion in $\eps\partial^2\vp$. 
Higher orders  in $\eps\partial^2\vp$ vanish as $\eps\to0$.
\begin{figure}
\includegraphics[width=3.6cm]{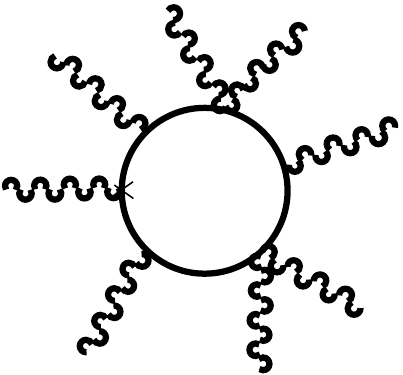} 
\caption{Graphical representation of the ratio of the determinants in \eq{R1} to first order
in $\eps\partial^2 \vp'$ denoted by a little cross.}
\label{anolog}
\end{figure}
The corresponding diagrams are shown in Fig.~\ref{anolog}.
All of them are of the same order in $b_0^2$.
The combinatorics is as follows:
\bea
{\rm Fig.~\ref{anolog}}&=&  
-\frac 12\sum_{n =1}^\infty \int \left(1-\e^{\vp'} \right)^{n}\partial ^2 \vp' 
\int \frac{\d^2 k}{(2\pi)^2}\frac{\eps}{(1+\eps k^2)^{n+1}}\non
&=&-\frac 1{8\pi}\sum_{n=1}^\infty \int \frac{\left(1-\e^{\vp'}  \right)^{n}}n\partial ^2 \vp' 
=\frac 1{8\pi}\int (\partial_a \vp')^2,
\eea
reproducing \rf{aa1}. 
This cancellation of the higher order in $\vp'$  is of cause a consequence of diffeomorphism  invariance. 

What was actually calculated was the partition function for the following simple modification
 of the Gaussian model
\be
Z=\int {\cal D} \vp' \e^{-\frac 1{16\pi b^2_0} \int \partial_a\vp' \partial_a \vp'} 
R_1^{-1}
\ee
with the ratio of the determinants $R_1$ given by \eq{R1}.
Because of the universality argument we expect that it is equivalent to the partition
function with the action \rf{S}. %% that reproduces KPZ-DDK.

\section{Renormalization of the exponential\label{s:renorm}}

\begin{figure}
\includegraphics[width=8cm]{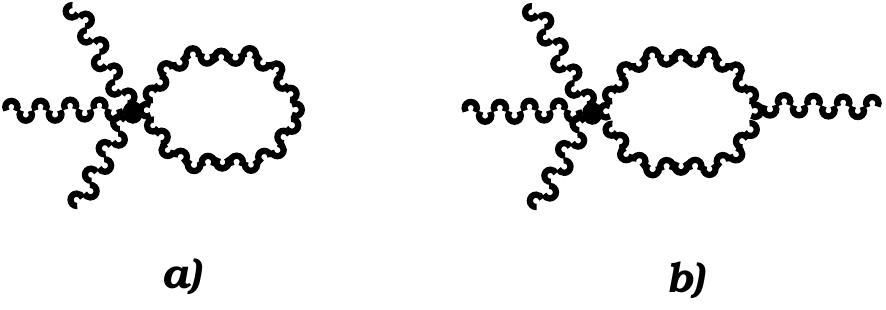}
\caption{One-loop renormalization of $\e^{\vp}$ whose position is denoted by the dot. 
The wavy lines represent $\vp$.}
\label{fi:fig3}
\end{figure}

While \eq{DDK1} comes from the requirement for the background 
$\hat \vp$ to disappear in the kinetic part
of the action after the shift \rf{vshift}, \eq{DDK2} follows from the independence of
$\int \e^{\alpha\vp}$
on $\hat \vp$. The term $-\alpha^2 b^2$ on the heft-hand side of \eq{DDK2} results from
the renormalization of $\e^{\alpha\vp}$ driven at one loop by the diagram in Fig.~\ref{fi:fig3}a 
which gives the logarithmic divergence
\be
{\rm Fig.~\ref{fi:fig3}a}=
 %%\frac{\vp^{n-2}}{2(n-2)!}
 \frac{\e^{\alpha\vp}}2\times
8\pi \alpha^2 b^2 \int \frac {\d^2 k}{(2\pi)^2} \frac 1{(k^2+\eps k^4 +\ldots)}= 
%%%\frac{\vp^{n-2}}{(n-2)!}
\e^{\alpha\vp} \left(\alpha^2b^2 \log \frac 1\eps +{\rm IR~ divergent}\right).
\label{56}
\ee
%%Superficially it does not depend on $\vp$ but 
Recalling that the world-sheet cutoff
\be
\eps = \frac {a^2}\brho \e^{-\alpha\vp},
\label{invaar}
\ee
where $a^2$ is an invariant cutoff, we find
\be
{\rm Fig.~\ref{fi:fig3}a}=\e^{\alpha\vp} \alpha^3 b^2 \vp.
\label{58}
\ee
Exponentiating the one-loop contribution, we reproduce the second term 
on the heft-hand side of \eq{DDK2}.

Now we have at one loop additionally the diagram in Fig.~\ref{fi:fig3}b which  for the
action \rf{S} contributes the same value as \rf{58} expanded in $b_0^2$
\be
{\rm Fig.~\ref{fi:fig3}b}=%%\frac{\vp^{n-2}}{2(n-2)!}
\frac {\e^\vp}2
\times 8\pi b_0^2 \vp \int \frac {\d^2 k}{(2\pi)^2} 
\frac {\eps k^4}{(k^2+\eps k^4 )^2}=  \e^\vp %%%\frac{\vp^{n-2}}{(n-2)!}
b^2_0 \vp .
\label{vvv}
\ee

At this point one may wonder about the term $\partial^2_a \vp$ which can also appear
from the diagram in Fig.~\ref{fi:fig3}b.
But it comes multiplied by the integral
\be
\partial^2_a \vp\int \frac {\d^2 k}{(2\pi)^2} 
\frac {\eps k^2}{(k^2+\eps k^4 )^2} \sim \eps\partial^2_a \vp
\label{72}
\ee
which is  negligible as $\eps\to0$.

It is remarkably simple to show that both \rf{58} and
\rf{vvv} are indeed universal. For \rf{58} it is obvious with logarithmic accuracy.
Given the general action \rf{Sgen},
\eq{vvv} is modified as
\be
{\rm Fig.~\ref{fi:fig3}b}=
\frac {\e^\vp}2
\times 8\pi b^2_0 \vp \int \frac {\d^2 k}{(2\pi)^2} 
\frac {\eps k^4 F'(\eps k^2)}{[k^2F(\eps k^2]^2}=  
\frac 1{F(0)}\e^\vp b^2_0 \vp =  \e^\vp b^2_0 \vp .
\label{vvvgen}
\ee

\begin{figure}
\includegraphics[width=8cm]{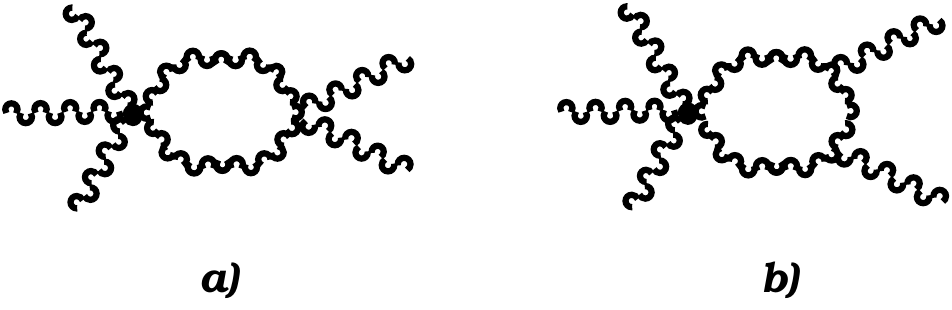} 
\caption{Cancellation of diagrams with two additional lines.}
\label{fi:fig4}
\end{figure}

A natural question arises as to the diagram with two additional lines depicted in
Fig.~\ref{fi:fig4}a whose contribution does not vanish and equals
\be
{\rm Fig.~\ref{fi:fig4}a}=
-\frac {\e^\vp}2
\times 8\pi b^2_0 \frac{\vp^2}2 \int \frac {\d^2 k}{(2\pi)^2} 
\frac {\eps k^4}{(k^2+\eps k^4)^2}=-\frac12 \e^\vp b^2_0 {\vp^2}  .
\ee
It mutually cancels with the diagram in Fig.~\ref{fi:fig4}b which has two vertices but
is of the same order in $b_0^2$:
\be
{\rm Fig.~\ref{fi:fig4}b}=
\frac {\e^\vp}2
\times 8\pi b^2_0 {\vp^2} \int \frac {\d^2 k}{(2\pi)^2} 
\frac {\eps^2 k^8}{(k^2+\eps k^4)^3}
=\frac12 \e^\vp b^2_0 {\vp^2}  .
\ee
This cancellation is again a consequence of diffeomorphism invariance and holds for
all additional powers of $\vp$.

The sum of Eqs.~\rf{58} and \rf{vvv} implies 
\be
\alpha=1+2b_0^2+{\cal O}(b_0^4)
\label{aaaa}
\ee 
which have the sense of finite renormalization of (the exponent in) the metric tensor.
Together with \eq{bbbb} it remarkably gives \eq{remark},
exactly reproducing \eq{b2DDK} of DDK to the given order.

\section{The Nambu-Goto string and simplified model\label{s:simp}}

%%\subsection{Divergent part}

To manage the Nambu-Goto action we proceed in the standard way, introducing the
(imaginary) Lagrange multiplier $\lambda^{ab}$ and an independent metric tensor
$\rho_{ab}$.
Path integrating over $X^\mu$,  their regulators and ghosts associated with 
fixing the conformal
gauge $\rho_{ab}=\rho \hat g_{ab}$ with  $\hat g_{ab}$
being  a fiducial world-sheet metric, we obtain for the divergent part of the effective
 action~\cite{AM16}
\be
S_{\rm div}=\int %\d^2 \omega 
\left[ \frac{K_0}2 \lambda^{ab} \partial_a X_{\rm cl}\cdot \partial_b X_{\rm cl}+
K_0  \rho\left(\sqrt{\det \hat g} -\frac12  \lambda^{ab} \hat g_{ab}\right) 
+\sqrt{\det \hat g} \left(-
\frac {d \Lambda^2 \rho}{2\sqrt{\det{\lambda}}}
+ \Lambda^2   \rho  \right)\right]. 
\label{cla}
\ee
Here $\Lambda$ is a UV cutoff, $K_0 \sim \Lambda^2$ is the bare string tension
and $X_{\rm cl}^\mu$ accounts for the boundary conditions imposed on the world-sheet,
{\it e.g.}\/ a long cylinder or torus..

Given  $\hat g_{ab}$, let us split the Lagrange multiplier
$\lambda^{ab}$ into the parts parallel and  orthogonal 
to $\hat g_{ab}$
\be
\lambda^{ab}=\lambda \sqrt{\det \hat g}\,\hat g^{ab}+\lambda_\perp^{ab}, \qquad  
\lambda_\perp^{ab} \hat g_{ab}=0 ,
\label{split}
\ee
where $\lambda$ is a scalar and $\lambda_\perp^{ab}$ is the orthogonal part.
We shall expand near  the minimum, substituting
$\lambda^{ab}=\blambda \sqrt{\det \hat g}\, \hat g^{ab}+\delta\lambda^{ab}$,
$\rho=\brho+\delta\rho$, so  the terms
linear in the fluctuations $\delta\lambda^{ab}$ and $\delta \rho$ will vanish in
the effective action.

It is convenient to work with complex coordinates $z=\omega^1+\i \omega^2$ and
$\bar z=\omega^1-\i \omega^2$ when 
\be
\lambda^{z\bz}=\lambda^{11}+\lambda^{22}, \quad
\lambda^{zz}=\lambda^{11}-\lambda^{22}+2\i \lambda^{12}.
\ee
Expanding in fluctuations about the minimum, we have for $\hat g_{ab}=\delta_{ab}$ 
the quadratic  in $\delta \lambda^{ab}$ part of the action
\be
S_{\rm div}^{(2)}=\int %\d^2 \omega  %%\sqrt{\det g}
\left[ -\Big(K_0-\frac {d\Lambda^2}{2 \blambda^2}\Big) \frac{\delta\lambda^{z\bz} \delta \rho}2 
-\frac {d\Lambda^2(\brho+\delta \rho)}{8 \blambda^3} \Big( (\delta \lambda^{z\bz})^2+
\frac 12\delta \lambda^{zz}\delta \lambda^{\bz\bz} \Big) \right] .
\label{Seff2+3zz}
\ee

It is seen from the action \rf{Seff2+3zz} that $\delta \lambda^{ab}$ does not propagate to
the distances much larger than $1/\Lambda \sqrt{\brho}$. Only $\delta \rho$ propagates
to macroscopic distances, so that the variables $\lambda^{zz}$, $\lambda^{\bz\bz}$ and
$\lambda^{z\bz}$ become localized. 
This is why the Nambu-Goto string is expected~\cite{Pol87}
to be equivalent to the Polyakov string. We would like to reexamine this issue, having
in mind that the anomalies of the type $\Lambda^{-2}\times\Lambda^2$ discussed above
might give again a contribution. 
%%We therefore may ignore finite parts for the terms where $\Lambda^2$ is present.

%\subsection{$\boldsymbol{\lambda_\parallel}$ case}

Let us first consider the case when $\lambda^{ab}_\perp=0$ in \eq{split}, \ie $\lambda^{zz}=\lambda^{\bz\bz}=0$ and  
$\lambda^{z\bz}=2\lambda$. Accounting for the finite part, the action for 
$\delta\lambda=\lambda-\blambda$ 
and $\delta\rho=\brho(\e^\vp-1)$ reads
\be
S^{(2)}=\int %\d^2 \omega  %%\sqrt{\det g}
\left[ \frac{(26-d)}{96\pi} \partial_a \vp \partial_a \vp
 -\frac{d}{24\pi \blambda} \partial_a \vp \partial_a \delta\lambda
- K_R\brho \e^\vp {\delta\lambda }
-\frac {d\Lambda^2\brho\e^\vp}{2 \blambda^3} \delta \lambda^2  \right] .
\label{Seff2}
\ee
We have dropped here a finite term for $\delta\lambda^2$ because of the presence of
$\Lambda^2$, but  left it for the mixed term $\delta\rho\delta\lambda$ because
\be
K_R=K_0-\frac {d\Lambda^2}{2 \blambda^2}
\ee
is finite in the scaling regime~\cite{AM16}.

We can path integrate over $\delta\lambda$.  The resulting action for $\vp$ is given
by the substitution of $\delta\lambda$ in \rf{Seff2} by
\be
\delta\lambda = \frac{\blambda^3}{d\Lambda^2} 
\left(\frac{d}{24\pi \blambda} \Delta \vp-K_R
\right).
\ee
The result would be of the type of the action \rf{S}, so nothing new appears in comparison with
the Polyakov string.

%\subsection{{$\boldsymbol{\lambda_\perp}$} case}

A simplest model where we may expect a deviation from the Polyakov string is
the Nambu-Goto string with frozen variable $\lambda^{z\bz}$ which can be ignored, but
still remaining $\lambda^{zz}$ and $\lambda^{\bz\bz}$. 
We thus  concentrate on the simplified model
generated by the quadratic action
\be
{\cal S}^{(2)}= \int \left[\frac 1{4\pi b^2_0} \partial \vp \bar \partial \vp + d \nu \left(
\lambda^{zz} \partial ^2  \vp
+\lambda^{\bar z \bz} \bar \partial ^2 \vp\right)- d{\Lambda^2 \brho}
%%\e^{\vp} 
\lambda^{zz} \lambda^{\bar z \bz}\right],
\ee
where $\lambda^{ab}$ is imaginary and $\nu$ is a constant.
Covariantizing it gives %
%%\footnote{For a general $\rho_{ab}$ we are to substitute here
%%$\vp$ by $(-\Delta^{-1} R)$ and $\e^\vp$ by $(\det \rho_{ab})^{1/2}$.}
\bea
{\cal S}&=&\int\left[
 \frac 1{16\pi b^2_0}( \det{\hat g})^{1/2}  \hat g^{ab}\partial
_a \vp \partial _b\vp +\frac 1{8\pi b^2_0}( \det{\hat g})^{1/2}  \hat R \vp
+ d \nu\left( \lambda^{ab}
\nabla _a \partial_b \vp 
 -\frac12 \lambda^{ab}\hat g_{ab} \hat g^{cd} \nabla_c \partial_d \vp\right)\right. \non
&&\hspace*{7mm}\left.-2d {\Lambda^2 \brho}
\e^{\vp}  ( \det{\hat g})^{-1/2} 
\left(\hat g_{ac} \hat g_{bd} -
\frac 12\hat g_{ab} \hat g_{cd}\right)\lambda^{ab} \lambda^{cd } \right].
\label{sm}
\eea

For $\hat g_{ab}=\delta_{ab}$ \eq{sm} yields the action
\bea
{\cal S}&= &\int\left[\frac 1{4\pi b_0^2} \partial \vp \bar \partial \vp + d \nu \left(
\lambda^{zz} \nabla \partial \vp
+ \lambda^{\bar z \bz}\bar \nabla\bar\partial \vp \right)- d{\Lambda^2 \brho}
\e^{\vp} \lambda^{zz} \lambda^{\bar z \bz} \right]\non
&=& \int\left[\frac 1{4\pi b_0^2} \partial \vp \bar \partial \vp + d \nu \left[
\lambda^{zz} \left(\partial^2\vp  - (\partial \vp )^2 \right)
+ \lambda^{\bar z \bz}\left(\bar \partial^2\vp - (\bar \partial{\vp})^2\right) \right]- 
d{\Lambda^2 \brho}
\e^{\vp} \lambda^{zz} \lambda^{\bar z \bz}\right] .
\label{21}
\eea
%%where $\vp'=\partial \vp$ and $\bar {\vp'}=\bar\partial \vp$.
At the one-loop order we can expand in $\vp$ to get only cubic and quartic  interactions.
%\be
%{\cal L}= \frac 1{4\pi b_0^2} \partial \vp \bar \partial \vp 
%+ d \nu \left( \lambda^{zz} \partial^2 \vp
%+\lambda^{\bar z \bz} \bar \partial ^2 \vp\right)  -d \nu \left[\lambda^{zz} (\partial \vp)^2
%+\lambda^{\bar z \bz} (\bar \partial \vp)^2  \right]
%- d{\Lambda^2 \brho} \left({1+\vp} +\vp^2+\ldots \right)\lambda^{zz} \lambda^{\bar z \bz}.
%\label{22}
%\ee

An interesting question is what modification of the usual conformal transformation
\begin{subequations}
\bea
\delta \vp& =& \xi'(z) +\xi(z) \partial \vp,\label{cta}\\ \delta 
\lambda^{ab} &=&\xi(z) \partial \lambda^{ab}
\label{ctb}
\eea
\label{ct}
\end{subequations}
\!\!would be the symmetry of \rf{21}? The first and the last terms on the right-hand
side of \eq{21} are invariant under \rf{ct}, while the other terms transform as
\bea
\delta \left(\lambda^{zz} \nabla \partial \vp \right)&=  &   \lambda^{zz}
\left(  \xi'''-\xi''\partial\vp+\xi' \nabla\partial \vp  \right)  ,       \\
\delta \left(\lambda^{\bar z \bz}\bar \nabla\bar\partial \vp \right)&=&
-\lambda^{\bar z \bz}\xi' \bar\nabla\bar\partial \vp.
\eea
The action \rf{21} then  remains invariant if
\begin{subequations}
\bea
\delta \vp &= &\xi' + \xi\partial \vp, \label{26a} \\
\delta\lambda^{zz}&= &\xi \partial \lambda^{zz}-\frac { \e^{-\vp}}{\nu \Lambda^2\brho} 
\xi' \bar\nabla\bar\partial \vp,\label{26b}\\
\delta\lambda^{\bz\bz}&= &\xi \partial \lambda^{\bz\bz}+
\frac {\e^{-\vp}}{\nu \Lambda^2\brho}\left(  \xi'''-\xi''\partial\vp+
\xi' \nabla\partial \vp  \right)
\label{26c}
\eea
\label{26}
\end{subequations}
and we disregard in the action the terms of order $\Lambda^{-2}$ as is
justified for smooth fields.
There exists an analog of this transformation for the Nambu-Goto action as well.

For our simplified model $\lambda^{ab}$ enters the action \rf{21} quadratically, so
we can eliminate it using the equation of motion
\be
\lambda^{zz}= \frac {\nu} {\Lambda^2\brho} \e^{-\vp}\bar\nabla\bar\partial \vp,
\qquad
\lambda^{\bz\bz}= \frac {\nu}{ \Lambda^2\brho} \e^{-\vp}\nabla\partial \vp.
\label{27a}
\ee
Now the conformal  transformation \rf{26a} of $\vp$  generates the
 transformations \rf{26b} and \rf{26c} of $\lambda^{zz}$ and $\lambda^{\bz\bz}$.
Using \rf{27a} we write for the action \rf{21}
\bea
{\cal S}&= &\int\left[\frac 1{4\pi b_0^2} \partial \vp \bar \partial \vp + 
\frac {d \nu^2}{\Lambda^2 \brho} \e^{-\vp}(
\nabla \partial \vp)(\bar \nabla\bar\partial \vp) \right]\non
&=& \int \left[
\frac 1{4\pi b_0^2} \partial \vp \bar \partial \vp +\frac {d \nu^2}{\Lambda^2 \brho} \e^{-\vp} \left(\partial^2\vp  - (\partial \vp )^2 \right)
\left(\bar \partial^2\vp - (\bar \partial{\vp})^2\right)\right] .
\label{28}
\eea
The second term in the action \rf{28} modifies the Liouville action. It
is negligible for smooth fields $\vp$ but, as we have already seen, may
contribute in our case where typical virtual $\vp$ is not smooth. 

Notice the difference between the actions \rf{S} and \rf{28}.
To show it explicitly we can rewrite \rf{28}, integrating by parts, in an equivalent form
\be
{\cal S}  %%&= &\frac 1{4\pi b_0^2} \partial \vp \bar \partial \vp + 
%%\frac {d \nu^2}{\Lambda^2 \brho} \e^{-\vp}\left[
%%(\partial\bar\partial \vp)^2 +\partial \vp\bar\partial \vp \partial\bar\partial \vp\right] \non 
=\frac 1{4\pi b_0^2}\int \left\{\partial \vp \bar \partial \vp + 4\eps\e^{-\vp}\left[
(\partial\bar\partial \vp)^2 +\partial \vp\bar\partial \vp \partial\bar\partial \vp
\right]\right\}, \qquad \eps=\frac{\pi d \nu^2b_0^2}{\Lambda^2\brho}
\label{equi}
\ee
by using the identity
\bea
\e^{-\vp} \left(\partial^2\vp  - (\partial \vp )^2 \right)
\left(\bar \partial^2\vp - (\bar \partial{\vp})^2\right)&=&
\e^{-\vp}\left[
(\partial\bar\partial \vp)^2 +\partial \vp\bar\partial \vp \partial\bar\partial \vp
\right]  \non &&+\partial \left[\e^{-\vp} \partial \vp\left(\bar \partial^2\vp - (\bar \partial{\vp})^2\right)\right]-\bar\partial (\e^{-\vp} \partial \vp \partial\bar \partial \vp).
\eea
%It is possible to further rewrite the action \rf{equi} using yet another identity
%\be
%\e^{-\vp}\partial \vp\bar\partial \vp \partial\bar\partial \vp
%= \e^{-\vp}\left[ (\partial \bar\partial \vp)^2+
%\bar\partial \vp \partial^2\bar\partial \vp \right] -\partial (
%\e^{-\vp}\bar\partial\vp \partial \bar\partial \vp).
%\ee

The additional terms in the action \rf{equi} are transformed under the
infinitesimal conformal transformation \rf{cta} as
\begin{subequations}
\bea
\delta \!\int \e^{-\vp}
(\partial\bar\partial \vp)^2 &= &0, \\
\delta \int \! \e^{-\vp}
\partial \vp\bar\partial \vp \partial\bar\partial \vp &= &\int \!\e^{-\vp} \xi'' 
\bar \partial \vp \partial\bar \partial \vp .
\label{33b}
\eea
\end{subequations}
Treating $\eps$ as a coupling constant, the change \rf{33b} can be compensated to
order $\eps$ by the following modification of \rf{cta}
\be
\delta_\xi\vp = \xi'+ \xi \partial \vp +\eps \xi'' \e^{-\vp} \bar\partial \vp 
+{\cal O}(\eps^2) .
\label{mct}
\ee
The action \rf{equi} then remains invariant under \rf{mct} to order $\eps$.

Under the transformation \rf{mct} we obtain
\be
\delta_\xi \e^{\vp} = \partial \left(\xi \e^{\vp} \right) + \eps \xi'' \bar \partial \vp +{\cal O}(\eps^2).
\ee
The additional term is the derivative with respect to $\bar z$, thus preserving the invariance of the volume
\be
\delta_\xi \int \!\e^{\vp} =0 .
\ee
A very nice property of the modified conformal transformation \rf{mct} is that it
preserves to order $\eps$ the commutation relation
\be
\delta_\xi \delta _\eta \e^\vp-\delta_\eta \delta _\xi \e^\vp =
\delta _\zeta \e^\vp, \qquad  \zeta= \xi \eta'-\xi'\eta.
\label{mcomm}
\ee
We thus expect the Virasoro algebra at the classical level.

%%\subsection{The Nambu-Goto case}

We believe that the action \rf{equi} of the simplified model captures certain characteristic 
features of the Nambu-Goto action. For the latter we have additionally a path integral 
over $\lambda^{z\bz}$ which was frozen in the simplified model. But this kind of path integration, which we discussed above,
can only modify the coefficients of the two terms with quartic derivatives
in \eq{equi} (as well as of the higher
terms dropped there). We thus may expect that path integrating first over $X^\mu$ and 
then over $\lambda^{ab}$  we obtain for the Nambu-Goto string to the first order in $\eps$
the action
\be
{\cal S} =\frac 1{4\pi b_0^2}\int \Big\{\partial \vp \bar \partial \vp + 4\eps\e^{-\vp}\left[
(\partial\bar\partial \vp)^2 +G\partial \vp\bar\partial \vp \partial\bar\partial \vp
\right]\Big\}
\label{equiNG}
\ee
with a certain constant $G$. For the simplified model~\rf{21} we have $G=1$.
The difference from the action \rf{S}, resulting from the Polyakov string, is by the presence
of the term with $G$.
Higher-order in $\eps$ terms are also possible but will not
change the result of the next section if universality holds \'a la Sect.~\ref{s:univ}.

\section{New anomaly for the Nambu-Goto string\label{s:NGnew}}

It is possible to compute the contribution of the term with $G$ in \rf{equiNG}
to the renormalization of the one-loop effective action. There is no contribution like $G^2$
because of the structure of the derivatives, so we have only a mixed contribution
\be
{\rm Fig.~\ref{fi:tad2}a}
=  -\frac 12 G\int (\partial_a\vp)^2
%%\frac{\d^2 p}{(2\pi)^2}\vp^2(p) \,
\int \frac{\d^2 k}{(2\pi)^2}\frac{\eps k^2 \eps k^4}{(k^2+\eps k^4)^2}
\label{pp1G}
\ee
and the tadpole
\be
{\rm Fig.~\ref{fi:tad2}b}
= \frac 12  G\int (\partial_a\vp)^2
%%\frac{\d^2 p}{(2\pi)^2}\vp^2(p) \,
\int \frac{\d^2 k}{(2\pi)^2}\frac{\eps k^2}{(k^2+\eps k^4)}
\label{pp2G}
\ee
which is like \rf{pp1} and \rf{pp2}. Each of the two is logarithmically divergent, but
the sum of \rf{pp1G} and \rf{pp2G} is finite and equals 
\be
{\rm Fig.~\ref{fi:tad2}a+Fig.~\ref{fi:tad2}b}  = \frac G{8\pi} \int (\partial_a\vp)^2.\label{ppfinG}
\ee

Summing with the previous result \rf{bbbb} for the action \rf{S}, we obtain
\be
\frac1{b^2}=\frac1{b_0^2} -\frac 16+2 +2G+{\cal O}(b_0^2).
\label{bbbbG}
\ee

As far as the renormalization of the metric tensor is concerned, there is obviously no
contribution from the term with $G$ to it because of the structure of the derivatives.
It is suppressed by $\eps$ similarly to \eq{72}. Therefore $\alpha$ is not affected by $G$
and is still given by \eq{aaaa}. We have thus observed a discrepancy with DDK for $G\neq0$.

Following the consideration in Sect.~\ref{Jacob}, we can 
discuss how the change of the action~\rf{equiNG} can be induced by
a field redefinition. Let us perform the change
\be
\vp\to \vp'= \vp -\frac\eps2 \e^{-\vp} \left( \partial_a^2 \vp+ G
\partial_a \vp \partial_a \vp
\right)+{\cal O}(\eps^2)
\label{che1}
\ee
for which
\be
\int \partial_a\vp \partial_a \vp \to \int \partial_a\vp' \partial_a \vp'=
\int \left\{\partial_a\vp \partial_a \vp +\eps \e^{-\vp}
\left[ (\partial_a^2 \vp)^2 +G\partial_a 
 \vp \partial_a\vp \partial_b^2 \vp
\right] \right\} +{\cal O}{(\eps^2)}.
\label{che2}
\ee
The change \rf{che1} produces the determinant of a nontrivial operator, so we
substitute $R_1$ given by \eq{R1} by
\be
R_1=
\frac{\det \left( \e^{\vp'}- \eps\partial_a^2  +\eps \partial_a^2\vp'/2+
G\eps\partial_a \vp'\partial_a - G\eps\partial_a \vp'\partial_a \vp'/2\right)}
{\det \left( \e^{\vp'} -\eps \partial_a ^2\right)}.
\label{rdet}
\ee
As before, we would naively expect that \rf{rdet} tends to 1 when $\eps\to0$, but
 it produces in fact the same anomaly as the sum of \rf{ppfin} and \rf{ppfinG}.

\section{Conclusion\label{s:conclu}}

The outcome of the calculation of this paper is twofold. First at all it was shown that straightforward 
computation for the Liouville theory with the action~\rf{S} modified by adding 
the higher-derivative term emerged for the Polyakov string, 
which is classically negligible for smooth fields as $\eps\to0$ but quantumly produces uncertainties like
powers of $\eps \times \eps^{-1}$ owing to the attendant interaction, 
agrees at one loop with KPZ-DDK. The result of doing these uncertainties
turned out to be universal and not dependent on the form of possible higher terms like it occurs
for quantum anomalies. To my knowledge it is for the first time
when KPZ-DDK is reproduced by direct computations without any assumptions.

Secondly, the situation with the Numbu-Goto string is different. Now the second term 
in the square brackets in \eq{equiNG} emerges additionally to the first one which was
the only one for the Polyakov string. It
is also important and changes the results. 
This seemingly implies that the Nambu-Goto and Polyakov
strings may be not equivalent at one loop.
For a more definite conclusion we need more calculations, in particular that of $q$.
If correct this might explain the deviation from the Alvarez-Arvis
string spectrum discovered in \cite{DFG12,AK13,Hel14} for an open outstretched Nambu-Goto
string, which is hard to understand for the Polyakov string.
It was one of my original motivations for this work.
%%%%%between the points separated by the distance $R$. 
The Alvarez-Arvis formula is associated in our approach with the ground state, 
while the observed deviation corresponds to one loop in our approach.

\subsection*{Acknowledgement}

This work was supported by the Russian Science Foundation (Grant No.20-12-00195).

\end{document}